\documentclass[twocolumn,twoside,11pt]{article}
\usepackage{fullpage,imr,jeffe,url}
\usepackage{graphicx}

\newtheorem{lemma}{Lemma}
\newtheorem{theorem}[lemma]{Theorem}

\def\hp{\hat{p}}

\def\hq{\hat{q}}

\def\hr{\hat{r}}
\def\hM{\hat{M}}
\def\grad{\nabla}
\def\subH{_{\!_H}\!}
\def\subZ{_{\!_Z}\!}
\def\subF{_{\!_F}\!}
\def\pH{p\subH}
\def\pF{p\subF}
\def\pZ{p\subZ}
\def\gradH{\nabla_{\!\!_H}}
\def\Line#1{\!\overlrarrow{\,\vphantom{x}#1\,}\!}

\begin{document}

\pagestyle{myheadings}
\markboth{Jeff Erickson, Damrong Guoy, John M. Sullivan, and Alper \Ungor}
	 {Building Space-Time Meshes over Arbitrary Spatial Domains}

\title{Building Space-Time Meshes\\over Arbitrary Spatial Domains}
\author{Jeff Erickson$^*$
	\and
	Damrong Guoy$^\dagger$
	\and
	John M. Sullivan$^\ddagger$
	\and
	Alper \Ungor$^{*\dagger}$
}

\date{$^*$Department of Computer Science\\
	$^\dagger$Computational Science and Engineering Program\\
	$^\ddagger$Department of Mathematics\\[1.5ex]
	Center for Process Simulation and Design\\
	University of Illinois at Urbana-Champaign\\
	\{jeffe,guoy,jms,ungor\}@uiuc.edu
}

\abstract{We present an algorithm to construct meshes suitable for 
	space-time discontinuous Galerkin finite-element methods.  Our
	method generalizes and improves the `Tent Pitcher' algorithm
	of \Ungor\ and Sheffer.  Given an arbitrary simplicially
	meshed domain~$X$ of any dimension and a time interval
	$[0,T]$, our algorithm builds a simplicial mesh of the
	space-time domain $X\times[0,T]$, in constant time per
	element.  Our algorithm avoids the limitations of previous
	methods by carefully adapting the durations of space-time
	elements to the local quality and feature size of the
	underlying space mesh.}


\maketitle

\begin{figure*}
\centerline{\includegraphics[height=3in]{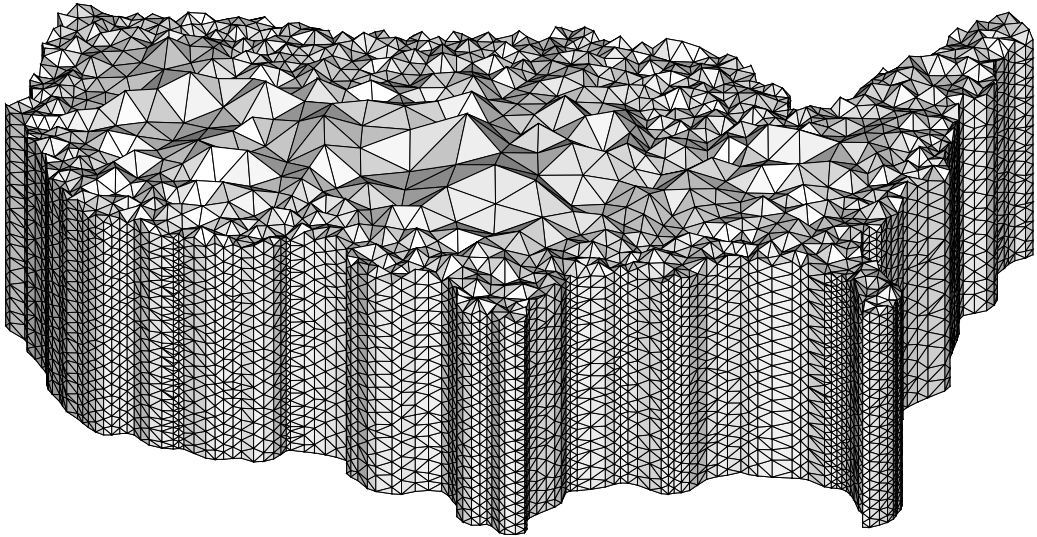}}
\caption{A space-time discontinuous Galerkin finite element mesh.}
\label{Fig:usa}
\end{figure*}


\section{Introduction}  

Many simulation problems consider the behavior of an object or region
of space over time.  The most common finite element methods for this
class of problem use a meshing procedure to discretize space, yielding
a system of ordinary differential equations in time.  A time-marching
or time-integration scheme is then used to advance the solution over a
series of fixed time steps.  In general, a distinct spatial mesh may
be required at each time step, due to the requirements of an adaptive
analysis scheme or to track a moving boundary or interface within the
domain.

A relatively new approach to such simulations suggests directly
meshing in space-time \cite{Lowrie98, Thompson94, YinASHT00}.  For
example, a four-dimensional space-time mesh would be required to
simulate an evolving three-dimensional domain.  Usually, the time
dimension is not treated in the same way as the spatial dimensions, in
part because it can be scaled independently.  Moreover, the numerical
methods that motivate our research impose additional geometric
constraints on the meshes to support a linear-time solution strategy.
Thus, traditional meshing techniques do not apply.

In this paper, we develop the first algorithm to build graded 
space-time meshes over arbitrary simplicially meshed domains in 
arbitrary dimensions.  Our algorithm does not impose a fixed global 
time step on the mesh; rather, the duration of each space-time element 
depends on the local feature size and quality of the underlying space 
mesh.  Our approach is a generalization of the `Tent Pitcher' 
algorithm of \Ungor\ and Sheffer \cite{UngorS01}, but avoids the 
restrictions of that method by imposing some additional constraints.  
Our algorithm builds space-time meshes in constant time per element.

The paper is organized as follows.  In Section~\ref{S:problem}, we 
formalize the space-time meshing problem and describe several previous 
results.  Section~\ref{S:front} explains the high-level advancing 
front strategy of our meshing algorithm.  In Sections~\ref{S:triangle} 
and~\ref{S:plane}, we develop our algorithm for building 
three-dimensional space-time meshes over triangulated planar domains.  
We generalize our algorithm to higher dimensions in 
Section~\ref{S:space}.  In Section \ref{S:output}, we describe our 
implementation and present some experimental results.  Finally, we 
conclude in Section~\ref{S:outro} by suggesting several directions for 
further research.


\section{Space-Time Discontinuous Galerkin Meshing}
\label{S:problem}

The formulation of our space-time meshing problem relies on the
notions of domain of influence and domain of dependence.  Imagine
dropping a pebble into a pond; over time, circular waves expand
outward from the point of impact.  These waves sweep out a cone in
space-time, called the domain of influence of the event.

More generally, we say that a point~$\hp$ in space-time \emph{depends
on} another point~$\hq$ if the salient physical parameters at~$\hp$
(temperature, pressure, stress, momentum, etc.\@) can depend on the
corresponding parameters at~$\hq$, that is, if changing the conditions
at $\hq$ could change the conditions at~$\hp$.  The \emph{domain of
influence} of~$\hp$ is the set of points that depend on~$\hp$;
symmetrically, the \emph{domain of dependence} is the set of points
that $\hp$ depends on.  At least infinitesimally, these domains can be
approximated by a pair of circular cones with common apex~$\hp$.  For
isotropic problems without material flow, this double cone can
described by a scalar \emph{wave speed} $c(\hp) \in \Real$, which
specifies how quickly the radius of the cones grows as a function of
time.  If the characteristic equations of the analysis problem are
linear and the material properties are homogeneous, the wave speed is
constant throughout the entire space-time domain; in this case, we can
choose an appropriate time scale so that ${c(\hp) = 1}$ everywhere.
For more general problems, the wave speed varies across space-time as
a function of other physical parameters, and may even be part of the
numerical solution.

These notions extend to finite element meshes in space-time.  We say 
that an element $\triangle$ in space-time depends on another 
element~$\triangle'$ if any point $\hp\in\triangle$ depends on any 
point $\hq\in\triangle'$.  This relation naturally defines a directed 
\emph{dependency graph} whose vertices are the elements of the mesh.  
Two elements in the mesh are \emph{coupled} if they lie on a common 
directed cycle in (the transitive closure of) the dependency graph.

\begin{figure}
\centerline{\includegraphics[height=1.5in]{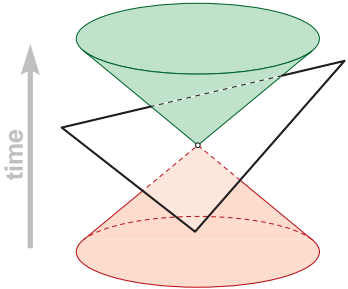}}
\caption{The cone constraint: Any boundary facet separates the domain 
of influence (above) from the domain of dependence (below).}
\label{Fig:conecon}
\end{figure}

Space-time discontinuous Galerkin (DG) methods have been proposed by 
Richter \cite{Richter94}, Lowrie \etal~\cite{Lowrie98}, and Yin 
\etal~\cite{YinASHT00} for solving systems of nonlinear hyperbolic 
partial differential equations.  These methods provide a linear-time 
element-by-element solution, avoiding the need to solve a large system 
of equations, provided no two elements in the underlying space-time 
mesh are coupled.  In particular, every pair of adjacent elements must 
satisfy the so-called \emph{cone constraint}: Any boundary facet 
between two neighboring elements separates the cone of influence from 
the cone of dependence of any point on the facet.  See 
Figure~\ref{Fig:conecon}.  Intuitively, if a boundary facet satisfies 
the cone constraint, information can only flow in one direction across 
that facet.  In a totally decoupled mesh, the dependency graph 
describes a partial order on the elements, and the solution can be 
computed by considering the elements one at a time according to any 
linear extension of this partial order.  Alternatively, the solutions 
within any set of incomparable elements can be computed in parallel.

Discontinuous Galerkin methods impose no \emph{a priori} restrictions
on the shape of the individual elements; mixed meshes with
tetrahedral, hexahedral, pyramidal, and other element shapes are
acceptable.  However, it is usually more convenient to work with very
simple convex elements such as simplices.  Experience indicates that
ill-conditioning is likely if the elements are non-convex, and
subdividing non-convex regions into simple convex elements is useful
for efficient integration.  (For further background on DG methods, we
refer the reader to the recent book edited by Cockburn, Karniadakis,
and Shu \cite{CockburnKS00}, which contains both a general survey
\cite{CockburnKS00a} and several papers describing space-time DG
methods and their applications.)

To construct an efficient mesh with convex elements, we have found
it preferable to relax the cone constraint in the following way.  
We construct a mesh of simplicial elements, but not all facets    
meet the cone constraint.  Instead, elements are grouped into     
patches (of bounded size).  The boundary facets between patches    
by definition satisfy the cone constraint, so patches are
partially ordered by dependence, and can be solved independently.

However, the internal facets between simplicial elements within
a patch may violate the cone constraint.  Thus, DG methods require
the elements within the patch to be solved simultaneously. 
Since each patch contains a constant number of elements,   
the system of equations within it has constant size, which 
implies that we can still solve the underlying numerical problem in
linear time by considering the patches one at a time.

Richter \cite{Richter94} observed that the dissipation of DG methods
increases as the slope of boundary facets decreases below the local
wave speed.  Thus, our goal is to construct an efficient simplicial
mesh, grouped into patches each containing few simplices, such that
the boundary facets of each patch are as close as possible to the cone
constraint without violating it.

\subsection*{Previous Results}

Most previous space-time meshing algorithms construct a single mesh
layer between two space-parallel planes and repeat this layer (or its
reflection) at regular intervals to fill the simulation domain.  The
exact construction method depends on the type of underlying space
mesh.  For example, given a structured quad space mesh, the space-time
meshing algorithm of Lowrie \etal~\cite{Lowrie98} constructs a layer
of pyramids and tetrahedra.  Similarly,
\Ungor~\etal~\cite{UngorHLSHT00, UngorSHT01} build a single layer of
tetrahedra and pyramids over an acute triangular mesh, and Sheffer
\etal~\cite{ShefferUTH00, UngorSHT01} describe an algorithm to build a
single layer of hexahedra over any (unstructured) quad mesh.  All such
layer-based approaches suffer from a global time step imposed by the
smallest element in the underlying space mesh.  This requirement
increases the number of elements in the mesh, making the DG method
less efficient; it also increases the numerical error of the solution,
since many internal facets must lie significantly below the constraint
cone.

A few recent algorithms do not impose a global time step, but instead 
allows the durations of space-time elements to depend on the size of 
the underlying elements of the ground mesh.  The first such 
algorithm, due to \Ungor~\etal~\cite{UngorSH00}, builds a triangular 
mesh for a $(1+1)$-dimensional space-time domain by intersecting the 
constraint cones at neighboring nodes.  This method does not easily 
generalize to higher dimensions.  The most general space-time meshing 
algorithm to date is the `Tent Pitcher' algorithm of \Ungor\ and 
Sheffer \cite{UngorS01}.  Given a simplicial space mesh in any fixed 
dimension, where every dihedral angle is strictly less than 
$90^\circ$, Tent Pitcher constructs a space-time mesh of arbitrary 
duration.  Moreover, if every dihedral angle in the space mesh is 
larger than some positive constant, each patch in the space-time mesh 
consists of a constant number of simplices.

Unfortunately, the acute simplicial meshes that Tent Pitcher requires
are difficult to construct, if not impossible, except in a few special
cases.  Bern \etal~\cite{BernEG94} describe two methods for building
an acute triangular mesh for an arbitrary planar point set, and
methods are known for special planar domains such as
triangles~\cite{Manheimer60}, squares~\cite{CassidyL80, Eppstein}, and
some classes of polygons \cite{HanganIZ00, Maehara00}.  However, no
method is known for general planar domains or even for point sets in
higher dimensions.  It is an open problem whether the cube has an
acute triangulation; see \cite{Ungor01} for recent related results.

\subsection*{New Results}

In this paper, we present a generalization of the Tent Pitcher 
algorithm that extends any simplicial space mesh in $\Real^d$, for any 
$d\ge 1$, into a space-time mesh of arbitrary duration.  Like the Tent 
Pitcher algorithm, our algorithm does not rely on a single global time 
step.  Our algorithm avoids the requirement of an acute ground mesh by 
carefully adapting the duration of space-time elements to the quality 
of the underlying simplices in the space mesh.


\section{The Advancing Front}
\label{S:front}

Our algorithm is designed as an advancing front procedure, which
alternately constructs a patch of the mesh and invokes a space-time
discontinuous Galerkin method to compute the required solution within
that patch.  To simplify the algorithm description, we assume that the
wave speed is constant throughout space-time; specifically, by
choosing an appropriate time scale, we will assume that $c(\hp) = 1$
everywhere.  Our algorithm can be easily adapted to handle changing
wave speeds, provided the wave speed at any point is a non-increasing
function of time.  We discuss the necessary changes for non-constant
wave speeds at the end of Section \ref{S:plane}.

The input to our algorithm is a simplicial \emph{ground mesh} $M$ of
some spatial domain $X \subset \Real^d$, with the appropriate initial
conditions stored at every element.  The advancing front $\hM$ is the
graph of a continuous \emph{time function} $t:X \to \Real$ whose
restriction to any element of the ground mesh is linear.  Any any
stage of our algorithm, each element of the front satisfies the cone
constraint $\norm{\grad t}\le 1$.  We will assume the initial time
function is constant, but more general initial conditions are also
permitted.

\begin{figure}[t]
\centering
\begin{tabular}{c}
	\includegraphics[height=1in]{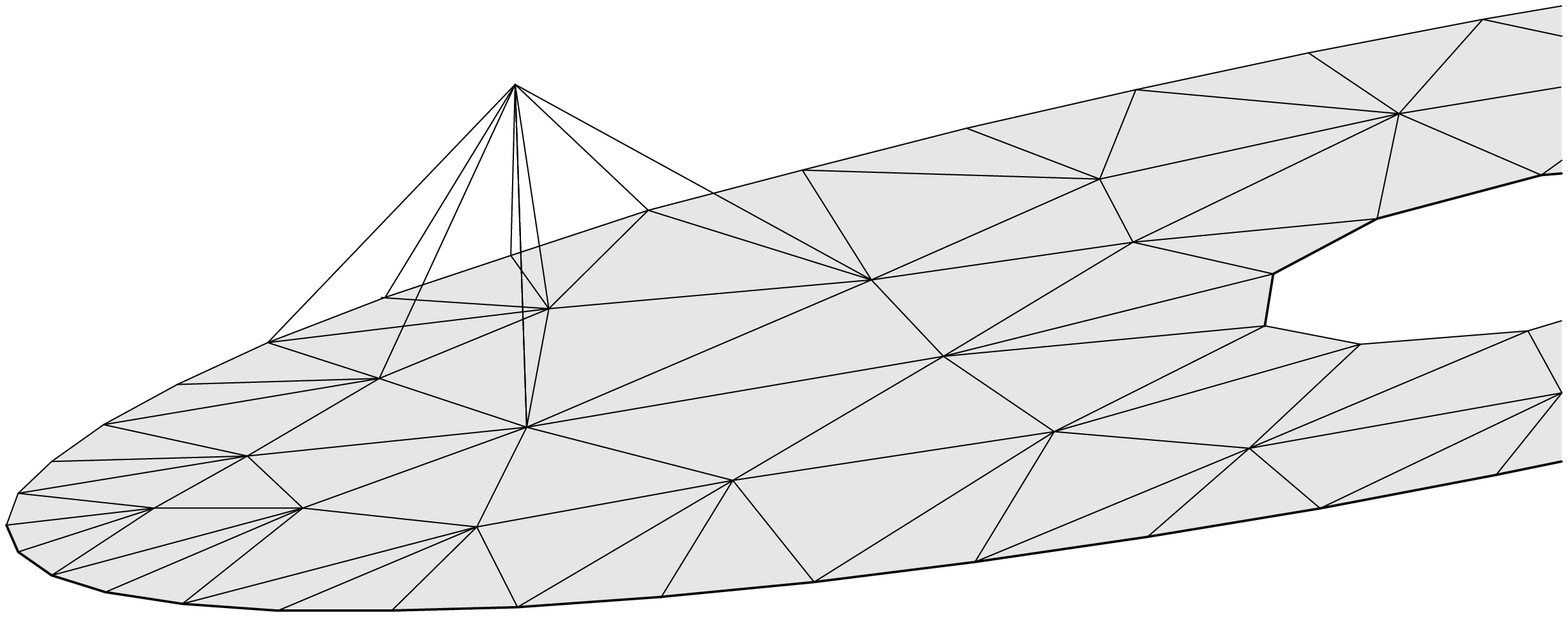}\\[2ex]
	\includegraphics[height=1in]{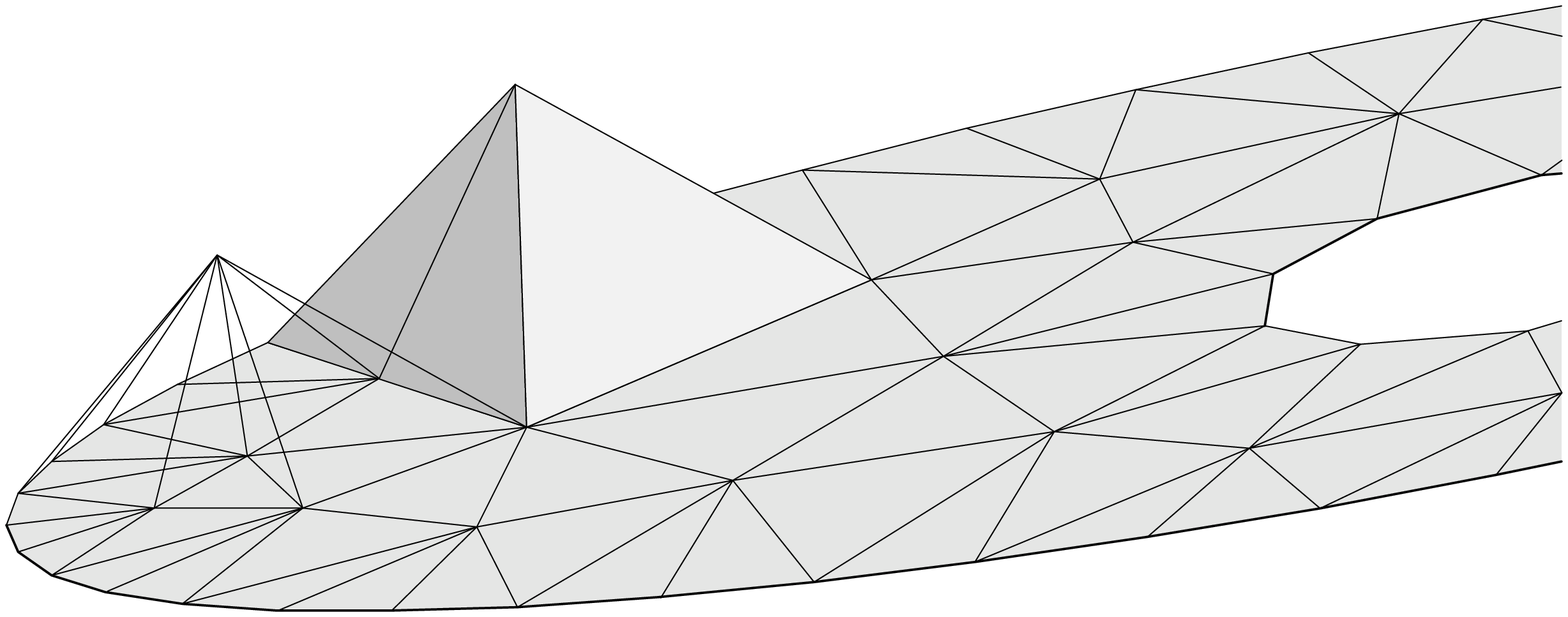}\\[2ex]
	\includegraphics[height=1in]{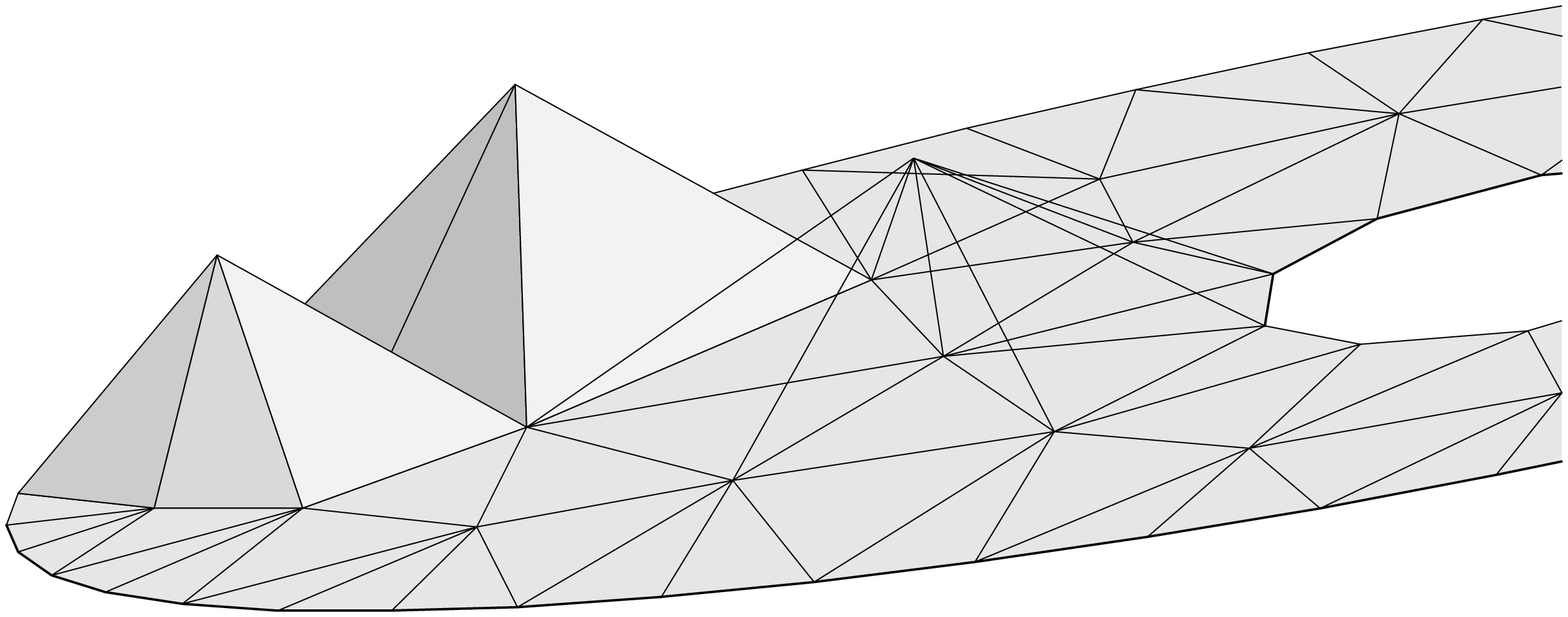}\\[2ex]
	\includegraphics[height=1in]{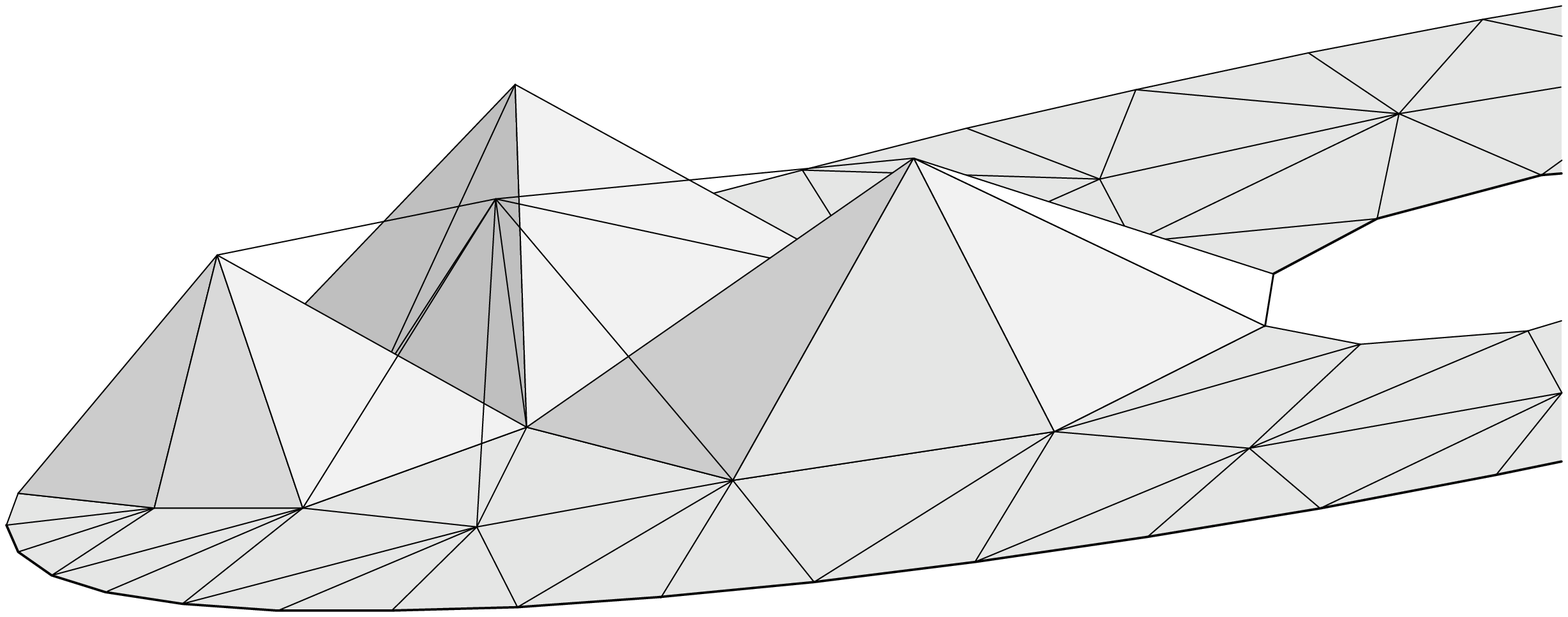}\\[2ex]
	\includegraphics[height=1in]{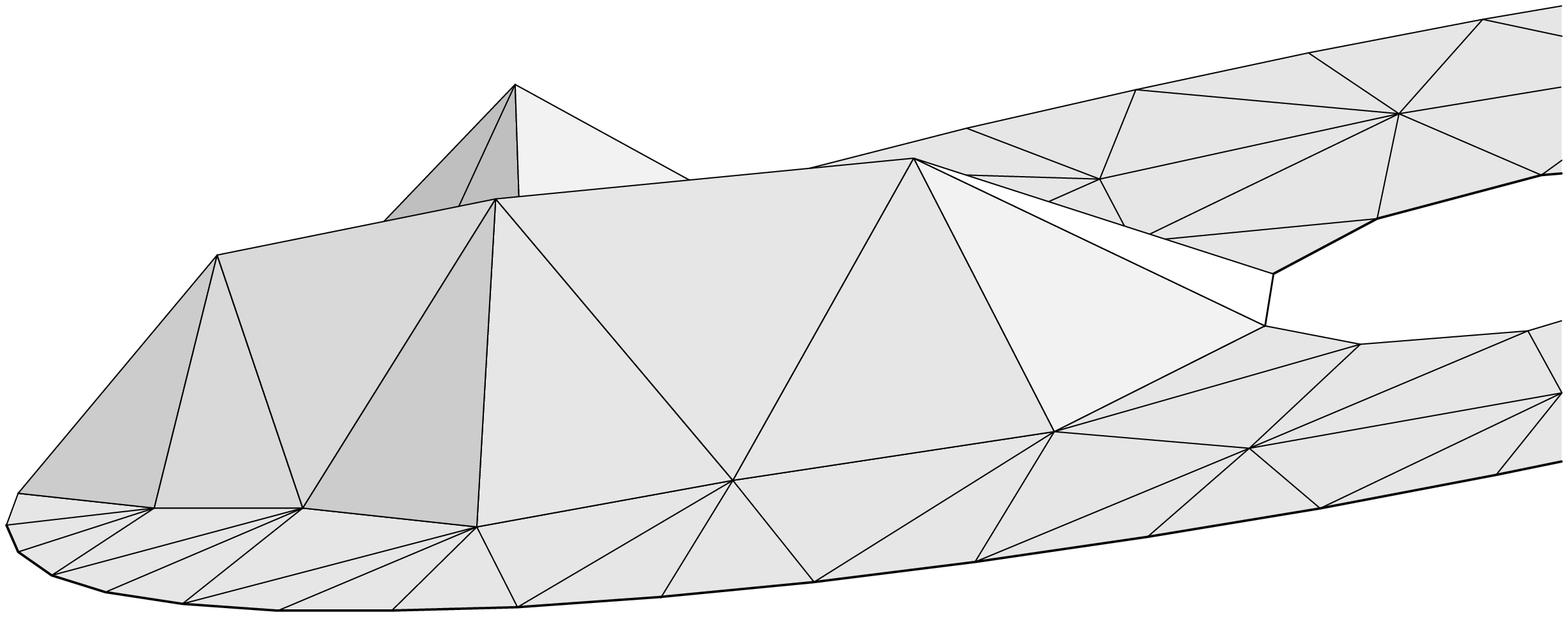}
\end{tabular}
\caption{Pitching a series of tents over a planar triangulation.}
\end{figure}

To advance the front, our algorithm chooses a vertex that is a local
minimum with respect to time, that is, a vertex $\hp = (p, t(p))$ such
that $t(p)\le t(q)$ for every neighboring vertex $\hq$.  (Initially,
\emph{every} vertex on the front is a local minimum.)  To obtain the
new front, this vertex is moved forward in time to a new point $\hp' =
(p, t'(p))$ with $t'(p) > t(p)$.  We call the volume between the the
old and new fronts a \emph{tent}.  The elements adjacent to $\hat{p}$
on the old front make up the \emph{inflow boundary} of the tent; the
corresponding elements on the new front comprise the patch's
\emph{outflow boundary}.  We decompose the tent into a patch of
simplicial elements, all containing the common edge $\hp\hp'$, and
pass this patch, along with the physical parameters at its inflow
boundary, to a DG solver.  The solver returns the physical parameters
for the outflow boundary, which we store for use as future inflow
data.  The solution parameters in the interior and inflow boundary of
the tent can than be written to a file (for later analysis or
visualization) and discarded.  This advancing step is repeated until
every node on the front passes some target time value.

If the front has several local minima, we could apply any number of
heuristics for choosing one; \Ungor\ and Sheffer outline several
possibilities~\cite{UngorS01}.  The correctness of our algorithm does
not depend on which local minimum is chosen.  In particular, if any
vertex has the same time value as one of its neighbors, we can break
the tie arbitrarily.  Our implementation computes the mesh in phases.
In each phase, we select a maximal independent set $S$ of local minima
and then lift each minimum in $S$, in some arbitrary order.  This
approach seems particularly amenable to parallelization, since the
minima in $S$ can be treated simultaneously by separate processors.


\section{Pitching Just One Triangle}
\label{S:triangle}

To complete the description of our algorithm, it remains only to
describe how to compute the new time value for each vertex to be
advanced, or less formally, how high to pitch each tent.  We first
consider the special case where the ground mesh consists of a single
triangle.  As we will show in the next section, this special case
embodies all the difficulties of space-time meshing over general
planar domains.

Let $p,q,r$ be three points in the plane.  At any stage of our
algorithm, the advancing front consists of a single triangle
$\triangle\hp\hq\hr$ whose vertices have time coordinates $t(p), t(q),
t(r)$.  Suppose without loss of generality that $t(p) < t(q) < t(r)$
and we want to advance $\hp$ forward in time.  We must choose the new
time value $t'(p)$ so that the resulting triangle $\triangle
\hp'\hq\hr$ satisfies the cone constraint $\norm{\grad t} \le 1$.

To simplify the derivation, suppose $q=(0,0)$ and $t(q)=0$.  The time
values $t'(p)$ and~$t(r)$ can then be written as $t'(p) = p \cdot
\grad t$ and $t(r) = r \cdot \grad t$, where $\grad t$ is the gradient
of the new time function.  We can write this gradient vector as
\[
	\grad t = \mu \bar{v} + \nu \bar{n},
\]
where $\bar{v}$ is the unit vector parallel to the vector $r$, and
$\bar{n}$ is the unit vector orthogonal to $\bar{v}$ with sign chosen
so that $\bar{n}\cdot p > 0$.  The vector $\mu\bar{v}$ is just the
gradient of the time function restricted to segment $qr$, so $\mu =
t(r) / \norm{r}$.  The cone constraint implies that $\norm{\grad t} =
\sqrt{\mu^2 + \nu^2} \le 1$ and therefore $\nu \le \sqrt{1 - \mu^2}$.
Thus, the cone constraint is equivalent to the following inequality:
\begin{align*}
	t'(p)
	&=
	p\cdot \grad t
\\	&=
	\mu p\cdot \bar{v} + \nu p\cdot \bar{n}
\\	&\le
	\mu p\cdot \bar{v} + \sqrt{1+\mu^2}\, p\cdot \bar{n}
\\	&=
	\frac{t(r)}{\norm{r}} \, p\cdot \bar{v}
	+
	\frac{\sqrt{\norm{r}^2 - t(r)^2}}{\norm{r}}\, p\cdot \bar{n}
\\	&=
	\frac{t(r)}{\norm{r}^2} \, p\cdot r +
	\frac{\sqrt{\norm{r}^2 - t(r)^2}}{\norm{r}^2}\, \abs{p\times r}
\end{align*}
Here, $p\times r$ denotes the two-dimensional cross product
$p_1r_2-p_2r_1$, which is just twice the signed area of
$\triangle{pqr}$.  To simplify the notation slightly, let $w_p$ denote
the distance from $p$ to $\smash{\Line{qr}}$, and define $w_q$ and
$w_r$ analogously:
\[
	w_p = \frac{2\abs{\triangle pqr}}{\norm{r-q}},
	\quad
	w_q = \frac{2\abs{\triangle pqr}}{\norm{p-r}},
	\quad
	w_r = \frac{2\abs{\triangle pqr}}{\norm{q-p}}.
\]
Then the previous inequality can be rewritten as
\begin{equation}
	t'(p) \le \frac{t(r)}{\norm{r}^2} \, p\cdot r +
	          \frac{\sqrt{\norm{r}^2 - t(r)^2}}{\norm{r}}\, w_p.
\label{eq:q=0}
\end{equation}
More generally, if $q\ne(0,0)$ and $t(q)\ne 0$, the cone constraint 
is equivalent to the following inequality.
\begin{equation}
	\fbox{\ensuremath{
	\begin{array}{@{}r@{}l@{}}
	t'(p) \le{} t(q)
	&{}+ \Frac{t(r) -t(q)}{\norm{r-q}^2} \, (p-q)\cdot (r-q)\\[3ex]
	&{}+ \Frac{\sqrt{\norm{r-q}^2 - (t(r)-t(q))^2}}
		  {\norm{r-q}}\,w_p
	\end{array}
	}}
\label{eq:cone}
\end{equation}
We have similar inequalities for every other ordered pair of vertices, 
limiting how far forward in time the lowest vertex can be moved past 
the middle vertex.  We will collectively refer to these six 
inequalities as the \emph{cone constraint}.

To ensure that our algorithm can create a mesh up to any desired time 
value, we must also maintain the following \emph{progress 
invariant}:
\begin{quote}
	The lowest vertex of $\triangle\hp\hq\hr$ can always be lifted 
	above the middle vertex without violating the cone constraint.
\end{quote}
This invariant holds trivially at the beginning of the algorithm, when 
$t(p)=t(q)=t(r)=0$.  Let us assume inductively that it holds at the 
moment we want to lift $\hp$.  \Ungor\ and Sheffer \cite{UngorS01} 
proved that if $\triangle pqr$ is acute, then satisfying the cone 
constraint automatically maintains this invariant, but for obtuse 
triangles, this is not enough.

To maintain our progress invariant, it suffices to ensure that in the
next step of the algorithm, the new lowest vertex $\hq$ can be lifted
above $\hr$ without violating the cone constraint.  In other words, if
we replace $t(q)$ with $t(r)$, the new triangle's slope must be
strictly less than $1$.  By substituting $t(r)$ for $t(q)$ in the cone
constraint \eqref{eq:cone} and making the inequality strict, we obtain
the following:
\begin{equation}
\fbox{\ensuremath{
	t'(p) < t(r) + w_p
}}
\label{eq:prog}
\end{equation}
We have similar inequalities for every other ordered pair of vertices, 
limiting how far forward in time the lowest vertex can be moved past 
the \emph{highest} vertex.  We will collectively refer to these six 
inequalities as the \emph{weak progress constraint}.

\begin{figure}[htb]
\centerline{\includegraphics[height=1.25in]{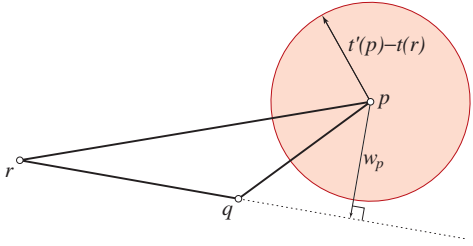}}
\caption{If the circle around $p$ does not touch the line through $q$ 
and $r$, then $\hq$ can be lifted above $\hr$ in the next step.}
\label{Fig:feasible}
\end{figure}

The weak progress constraint has a simple geometric interpretation,
which we can see by looking at the lifted triangle in space-time; see
Figure~\ref{Fig:feasible}.  Let $\Gamma$ be the cone of dependence of
the lifted point $\hat{p}'$; this cone intersects the plane $t = t(r)$
in a circle $\gamma$ of radius ${t'(p)-t(r)}$.  Any plane $\pi$
through $\hp'$ that satisfies the cone constraint is disjoint from
$\Gamma$; in particular, the intersection line of $\pi$ with the plane
$t=t(r)$ does not cross~$\gamma$.  Now let $\hq' = (q, t(r))$.  If the
plane $\hp'\hq'\hr$ satisfies the cone constraint, then the line
through~$\hq'$ and~$\hr$ does not cross~$\gamma$.  Thus, the progress
invariant holds after we lift~$\hp$ only if $t'(p) - t(r) < w_p$.

Our algorithm lifts $\hp$ to some point $\hp'$ that satisfies both the
cone constraint and the weak progress constraint, where $t'(p) >
t(q)$.  By the progress invariant, this does not violate the cone
constraint.  If $t'(p) \ge t(r)$, then the weak progress constraint
implies that the progress invariant still holds.  If $t'(p) < t(r)$,
then the progress invariant also still holds, because $t(r) - t'(p) <
t(r) - t(q)$.  Thus, by induction, the progress invariant is
maintained at every step of our algorithm.

Unfortunately, the weak progress constraint does not guarantee that we
can reach any desired time value; in principle, the advancing front
could converge to some finite limit.  To guarantee significant
progress at every step of the algorithm, we need a slightly stronger
constraint.  Our implementation uses the inequality
\begin{equation}
\fbox{\ensuremath{
	t'(p) \le t(r) + (1-\e)w_p
}}
\label{eq:mod}
\end{equation}
where $\e$ is a fixed constant in the range ${0 < \e \le 1/2}$.  We
have a similar inequality for every other ordered pair of vertices,
and we collectively refer to these six inequalities as the
\emph{progress constraint}.

With this stronger constraint in place, we have the following result.

\begin{lemma}
\label{L:progress}
If the cone constraint and progress constraint hold beforehand, we can
lift $\hat{p}$ at least $\e w_p$ above $\hat{q}$ without violating
either constraint.
\end{lemma}

\begin{proof}
Without loss of generality, assume that $q=(0,0)$ and $t(q) = 0$.  We
want to prove that setting $t'(p) = \e w_p$ does not violate the cone
constraint (in its simpler form \eqref{eq:q=0}) or the progress
constraint \eqref{eq:mod}.  Recall our assumption that $t(r) \ge t(q)
= 0$.   Because $\e \le 1/2$, we have
\[
	t'(p) = \e w_p \le (1-\e) w_p \le t(r) + (1-\e)w_p,
\]
so the progress constraint is satisfied.  The previous progress
constraint implies that $t(r) \le {(1-\e) w_r}$.  Because $\e>0$ and
$w_r \le \norm{r} = \norm{r-q}$, we have
\[	
	t(r)^2 \le (1-\e)^2w_r^2 \le (1-\e^2) \norm{r}^2,
\]
which implies that
\[
	\e \le \frac{\sqrt{\norm{r}^2 - t(r)^2}}{\norm r}.
\]
Finally, because $t(r)\ge 0$, we have
\begin{align*}
	t'(p) = \e w_p &\le \frac{\sqrt{\norm{r}^2 - t(r)^2}}{\norm r} w_p
\\	&\le  \frac{t(r)}{\norm{r}^2} \, p\cdot r +
		\frac{\sqrt{\norm{r}^2 - t(r)^2}}{\norm r} w_p.
\end{align*}
Thus, the cone constraint is also satisfied.
\end{proof}

\begin{theorem}
Given any three points $p,q,r \in \Real^2$, any real value $T>0$, and
any constant $0 < \e \le 1/2$, our algorithm generates a tetrahedral
mesh of the prism $\triangle pqr\times[0,T]$, where every internal
facet satisfies the cone constraint.  The number of tetrahedra is at
most $TP/2A\e$, where $P$ is the perimeter and $A$ is the area of
$\triangle pqr$.
\end{theorem}

\begin{proof}
Our algorithm repeatedly lifts the lowest vertex of $\triangle
\hp\hq\hr$ to the largest time value satisfying the cone constraint
\eqref{eq:cone}, the progress constraint \eqref{eq:mod}, and a
termination constraint $t\le T$.  Each time we lift a point, our
algorithm creates a new tetrahedron.  By Lemma \ref{L:progress}, a new
point becomes the lowest vertex, so the algorithm halts only when all
three vertices reach the target plane $t=T$.  Moreover, whenever $t(p)
\le t(q) \le t(r)$, the algorithm chooses a new time value $t'(p) \ge
t(q) + \e w_p \ge t(p) + \e w_p$, except possibly when $t'(p) = T$.
Thus, $\hp$ is lifted at most $T/\e w_p = T\norm{q-r}/2A\e$ times
before the algorithm terminates.
\end{proof}


\section{Arbitrary Planar Domains}
\label{S:plane}

We now extend our meshing algorithm to more complex planar domains.
The input is a triangular \emph{ground mesh}~$M$ of some planar
domain~$X$.  As we described in Section \ref{S:front}, our algorithm
maintains a polyhedral front~$\hM$ with a lifted vertex $\hp = (p,
t(p))$ for every vertex ${p\in M}$.  To advance the front, our
algorithm chooses a local minimum vertex~$\hp$ and lifts it to a new
point $\hp' = (p, t'(p))$.

The new time value $t'(p)$ is simply the largest value that satisfies
the cone constraints and progress constraints for every triangle in
the ground mesh that contains~$p$.  The chosen time value $t'(p)$ is
the value that would be chosen by at least one of these triangles in
isolation.  It follows that $\hp'$ is not a local minimum in the
modified front.  Moreover, by our earlier arguments, the progress
invariant is maintained in every triangle adjacent to~$p$.  It follows
immediately that our algorithm can generate meshes to any desired time
value.

Specifically, let $\omega_p$ denote the minimum distance from~$p$ to
$\smash{\Line{qr}}$, over all triangles $\triangle pqr$ in the ground
mesh.  Lemma \ref{L:progress} implies the following result.

\begin{theorem}
\label{Th:2dmesh}
Given any triangular mesh $M$ over any domain $X\subset \Real^2$, any
real value $T>0$, and any constant $0<\e\le 1/2$, our algorithm
generates a space-time mesh for the domain $X\times[0,T]$.  The number
of patches is at most $(T/\e) \sum_{p \in M} 1/\omega_p$, and number
of tetrahedra is at most $(6T/\e) \sum_{p \in M} 1/\omega_p$.
\end{theorem}

Our analysis of the number of patches and elements is conservative,
since it assumes that each step of the algorithm advances a vertex by
the minimum amount guaranteed by Lemma \ref{L:progress}.  We expect
most advances to be larger in practice, especially in areas of the
ground mesh without large angles.  Our experiments were consistent
with this intuition; see Section \ref{S:output}.

Most of the parameters of the cone constraint, and all of the 
parameters of the progress constraint, can be computed in advance from 
the ground mesh alone.  Thus, the time to compute each new time value 
$t'(p)$ is a small constant times the degree of $p$ in the ground 
mesh, and the overall time required to build the mesh is a small 
constant times the number of mesh elements.

\subsection*{Non-constant Wave Speeds}

Although we have described our algorithm under the assumption that the
wave function $c(\hp)$ is constant, this assumption is not necessary.
If elements of the ground mesh have different (but still constant)
wave speeds, our algorithm requires only trivial modifications.  The
situation fits well with discontinuous Galerkin methods, which compute
solutions with discontinuities at element boundaries.  If the wave
speed varies within a single element, even discontinuously, the only
necessary modification is to compute and use the maximum wave speed
over each entire element.  Similar modifications suffice if the wave
speed at any point in space can decrease over time.

If the mesh has only acute angles, the progress constraint is
redundant and arguments of \Ungor\ and Sheffer~\cite{UngorS01} imply
that our algorithm works even if the wave speed can increase over
time, as long as the wave speed is Lipschitz continuous.
Unfortunately, their analysis breaks down for obtuse meshes because of
the progress constraint, and indeed our algorithm can get stuck.  We
expect that a refinement of our progress constraint would allow for
increasing wave speeds, but further study is required.


\section{Higher Dimensions}
\label{S:space}

Our meshing algorithm extends in an inductive manner to simplicial
meshes in higher dimensions.  As in the two-dimensional case, it
suffices to consider the case where the ground mesh consists of a
single simplex $\triangle$ in $\Real^d$.  At each step of our
algorithm, we increase the time value of the lowest of the simplex's
$d+1$ vertices as much as possible so that the cone constraint
$\norm{\grad t} \le 1$ is satisfied and we can continue inductively as
far into the future as we like.

\begin{figure}[htb]
\centerline{\includegraphics[width=3.25in]{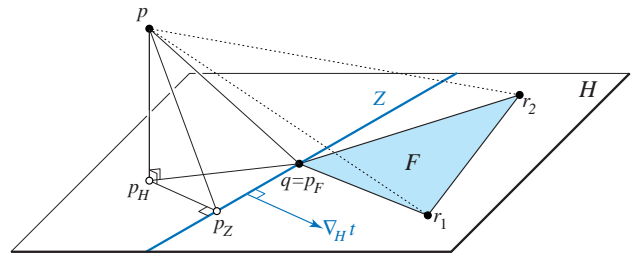}}
\caption{Defining the points $\pH$, $\pF$, and $\pZ$}
\label{Fig:project}
\end{figure}

Let $p, q, r_1, r_2, \dots, r_{d-1}$ denote the vertices of
$\triangle$ in increasing time order, breaking ties arbitrarily.  Our
goal is to lift $\hp$ above $\hq$ without violating the cone
constraint.  Let $F$ be the facet of $\triangle$ that excludes $p$,
and let $H$ be the hyperplane spanning~$F$.  Let $\pH$ be the
projection of $p$ onto $H$, and let $\pF$ be the closest point in $F$
to $p$.  Observe that $\angle p\pH\pF$ is a right angle.  See Figure
\ref{Fig:project}.  Let $\gradH t$ denote the gradient vector of the
time function restricted to $H$.  Finally, define
\[
	\sigma\subF = \frac{\norm{p-\pH}}{\norm{p-\pF}}.
\]
If $\pH$ lies inside $F$, then $\pH = \pF$ and $\sigma\subF = 1$; 
otherwise, $\pH \ne \pF$ and $\sigma\subF = \sin\angle \pH \pF p$.

The higher-dimensional analogue of the weak progress constraint is 
described by the following lemma.

\begin{lemma}
\label{L:facet}
If $\norm{\gradH t} < \sigma\subF$, then we can lift $\hp$ above $\hq$ 
without violating the cone constraint $\norm{\grad t}\le 1$.
\end{lemma}

\begin{proof}
Suppose $\norm{\gradH t} < \sigma\subF$.  Without loss of generality, 
assume that $q = (0, 0, \dots, 0)$ and $t(q) = 0$.  Let $\bar{n}$ be 
the unit normal vector of $H$ with $p\cdot \bar{n} > 0$,
\[
	\bar{n} = \frac{p-\pH}{\norm{p-\pH}}.
\]
Since the time function $t$ is linear, changing only $t(p)$ is 
equivalent to leaving $t$ fixed on the hyperplane $H$ and changing the 
directional derivative $\partial t/\partial \bar{n}$.  To prove the 
lemma, we show that setting
\begin{equation}
	\frac{\partial t}{\partial \bar{n}}
	=
	\cos \angle \pH\pF p
	=
	\frac{\norm{\pH-\pF}}{\norm{p-\pF}}
\label{eq:cos}
\end{equation}
gives us a new time function that satisfies the cone constraint with 
$t(p)>0$.

Let $Z$ be the set of points in $H$ where $t = 0$.  Since $t(q)=0$,
$Z$ is the $(d-2)$-flat orthogonal to $\gradH t$ that passes
through~$q$.  Moreover, because $t\ge 0$ everywhere in $F$, $Z$ is a
supporting $(d-2)$-flat of $F$.  Let $\pZ$ be the closest point in $Z$
to $p$ (or to $\pH$); this might be the same point as $\pF$, $\pH$,
or~$q$.  Observe that $\angle p\pH\pZ$ is a right angle.  See
Figure~\ref{Fig:project}.

We can express the time gradient $\grad t$ as follows:
\[
	\grad t = \gradH t + \frac{\partial t}{\partial \bar{n}} \bar{n}.
\]
Equation (\ref{eq:cos}) implies that
\[
	\grad t = \gradH t + \frac{\norm{\pH-\pF}}{\norm{p-\pF}} \bar{n}.
\]
Since these two components of $\grad t$ are orthogonal, we can 
express its length as follows.
\begin{align*}
	\norm{\grad t}^2
	&= 
	\norm{\gradH t}^2 + 
	\frac{\norm{\pH-\pF}^2}{\norm{p-\pF}^2}
\\	&< 
	\frac{\norm{p-\pH}^2}{\norm{p-\pF}^2} +
	\frac{\norm{\pH-\pF}^2}{\norm{p-\pF}^2}
	= 1
\end{align*}
So the new time function satisfies the cone constraint.

We can express the time value $t(p)$ as follows:
\begin{align*}
	t(p)
	&=
	t(\pH) + \norm{p-\pH} \frac{\partial t}{\partial \bar{n}}
\\	&=
	t(\pH) + \frac{\norm{p-\pH}\,\norm{\pH-\pF}}{\norm{p-\pF}}
\end{align*}
If $t(\pH)\ge 0$, then clearly $t(p) > 0$.  Suppose ${t(\pH) < 0}$.  
The vector $\pH-\pZ$ is orthogonal to~$Z$ and therefore anti-parallel 
to $\gradH t$.  Thus,
\begin{align*}
	t(\pH)
	&= \gradH t \cdot (\pH-\pZ)
\\	&= - \norm{\gradH t}\, \norm{\pH - \pZ}
\\	&\ge - \frac{\norm{p-\pH}\,\norm{\pH-\pZ}}{\norm{p-\pF}}
\\	&\ge - \frac{\norm{p-\pH}\,\norm{\pH-\pF}}{\norm{p-\pF}}.
\end{align*}
The last inequality follows from the fact that $\pH$~and~$F$ lie on
opposite sides of~$Z$, because $t(\pH) < 0$.  It now immediately
follows that $t(p) > 0$.
\end{proof}

As in the two-dimensional case, in order to guarantee that the
algorithm does not converge prematurely, we must strengthen this
constraint.  There are many effective ways to do this; the following
lemma describes one such method.

\begin{lemma}
\label{L:facet2}
For any $0<\e\le 1$, if $\norm{\gradH t} \le (1-\e)\sigma\subF$, then
we can lift $\hp$ at least $\e \norm{p-\pH}$ above $\hq$ without
violating the cone constraint $\norm{\grad t}\le 1$.
\end{lemma}

\begin{proof}
We modify the previous proof as follows.  We show that setting
\[
	\frac{\partial t}{\partial \bar{n}}
	=
	\e + (1-\e)\frac{\norm{\pH-\pF}}{\norm{p-\pF}}
\]
gives us a new time function satisfying the conditions of the lemma.
First we verify that the cone constraint is satisfied.
\begin{align*}
	\norm{\grad t}^2
	&\le
	\left((1-\e)\frac{\norm{p-\pH}}{\norm{p-\pF}}\right)^2 +
	\left(\e + (1-\e)\frac{\norm{\pH-\pF}}{\norm{p-\pF}}\right)^2
\\	&=
	1 + 2\e(1-\e) \left(\frac{\norm{\pH-\pF}}{\norm{p-\pF}} - 1\right)
\\	&\le 1
\end{align*}
(In fact, if $\pH\ne\pF$, then $\norm{\grad t} < 1$, which means we 
could lift $\hp$ even more.)

Next we verify that $t(p) \ge \e \norm{p - \pH}$.
\begin{align*}
	t(p)
	&=
	t(\pH) + \norm{p-\pH} \frac{\partial t}{\partial \bar{n}}
\\	&=
	t(\pH) + \e\norm{p-\pH}
		+ (1-\e)\frac{\norm{p-\pH}\,\norm{\pH-\pF}}
				{\norm{p-\pF}}
\\	&\ge
	t(\pH) + \e\norm{p-\pH}.
\end{align*}
If $t(\pH)\ge 0$, we are done.  Otherwise, as in the previous lemma, 
we have
\begin{align*}
	t(\pH)
	&\ge - \norm{\gradH t}\, \norm{\pH - \pF}
\\	&\ge -(1-\e)\frac{\norm{p-\pH}\,\norm{\pH-\pF}}{\norm{p-\pF}},
\end{align*}
which immediately implies that $t(p) \ge \e\norm{p-\pH}$, as claimed.
\end{proof}

An important insight is that we can view the simplex $\triangle$
simultaneously as a single $d$-dimensional simplex and as
$(d-1)$-dimensional boundary mesh.  Lemma~\ref{L:facet2} prescribes a
tighter cone constraint for every element in this boundary mesh.

Our algorithm proceeds as follows.  At each step, we lift the lowest
vertex of $\triangle$ by recursively applying the $(d-1)$-dimensional
algorithm; then, if necessary, we lower the newly-lifted vertex to
satisfy the global cone constraint $\norm{\grad t} \le 1$.  The base
case of the dimensional recursion is the two-dimensional algorithm in
the previous section.

This recursion imposes an upper bound on the length of the time
gradient within every face of $\triangle$ of dimension at least $1$.
In fact, a \naive\ recursive implementation would calculate $(d-k)!$
different constraints for each $k$-dimensional face.  A more careful
implementation would determine the strictest constraint for each face
in an initialization phase, so that each step of the algorithm only
needs to consider each face incident to the lifted vertex once.

For a $d$-dimensional ground mesh with more than one simplex, we apply 
precisely the same strategy as in the two-dimensional case.  At each 
step of the algorithm, we choose an arbitrary local minimum 
vertex~$\hp$, and lift it to the highest time point $\hp'$ allowed by 
all the simplices (of all dimensions) containing~$\hp$.  By our 
earlier arguments, $\hp'$ is not a local minimum of the modified 
front, which implies that our algorithm terminates only when all the 
vertices reach the target time value.


\section{Output Examples}
\label{S:output}

We have implemented our planar space-time meshing algorithm and tested
it on several different ground meshes.  Our implementation consists of
approximately 5000 lines of \Cplusplus\ code, about 800 of which
represent the actual space-time meshing algorithm; the remaining code
is a pre-existing library for manipulating and visualizing triangular
and tetrahedral meshes.

\begin{figure}[t]
\centering\footnotesize\sf
\begin{tabular}{@{}c@{}}
	\includegraphics[width=2.5in]{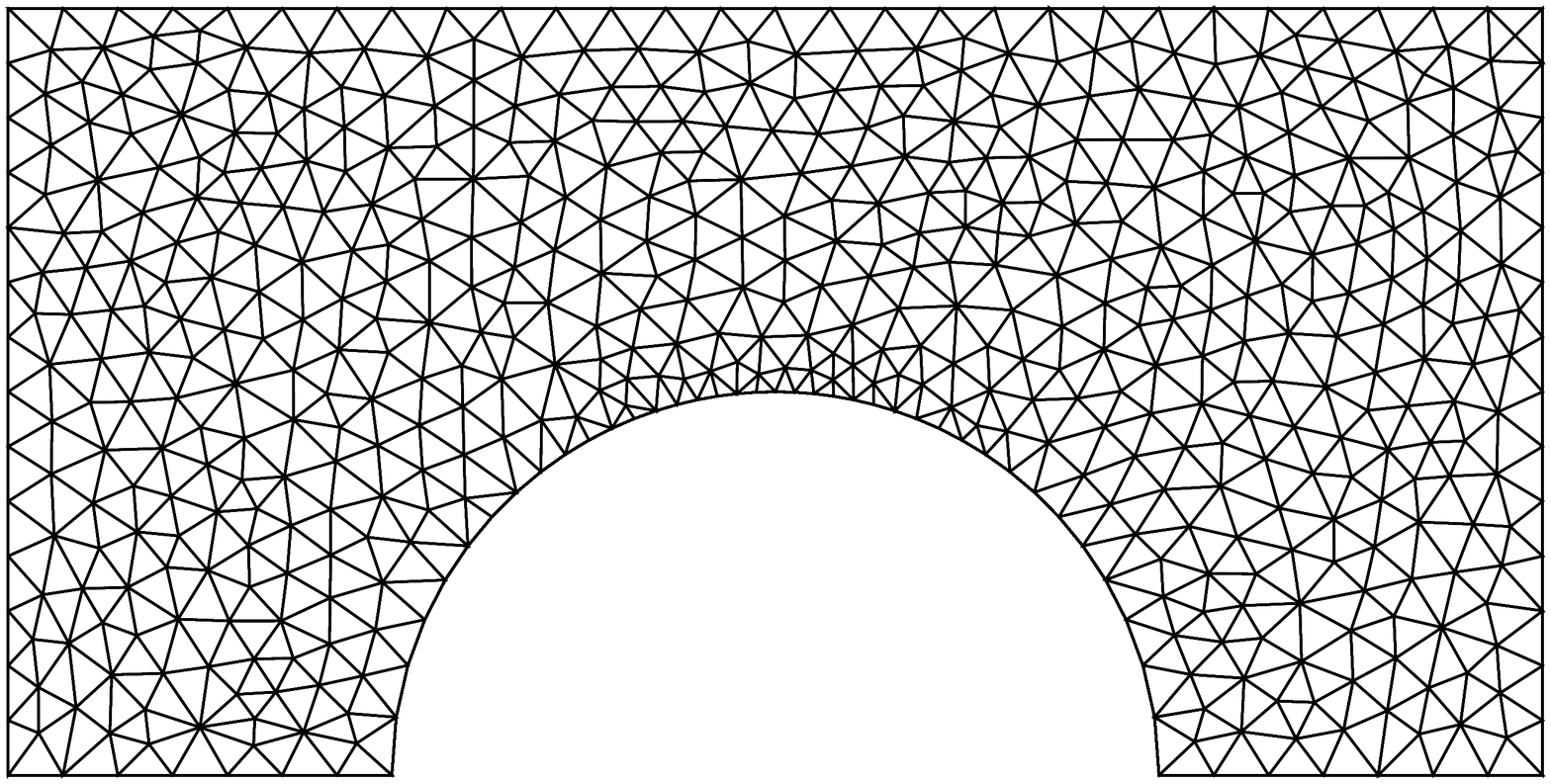} \\ (a) \\[2ex]
	\includegraphics[width=3in]{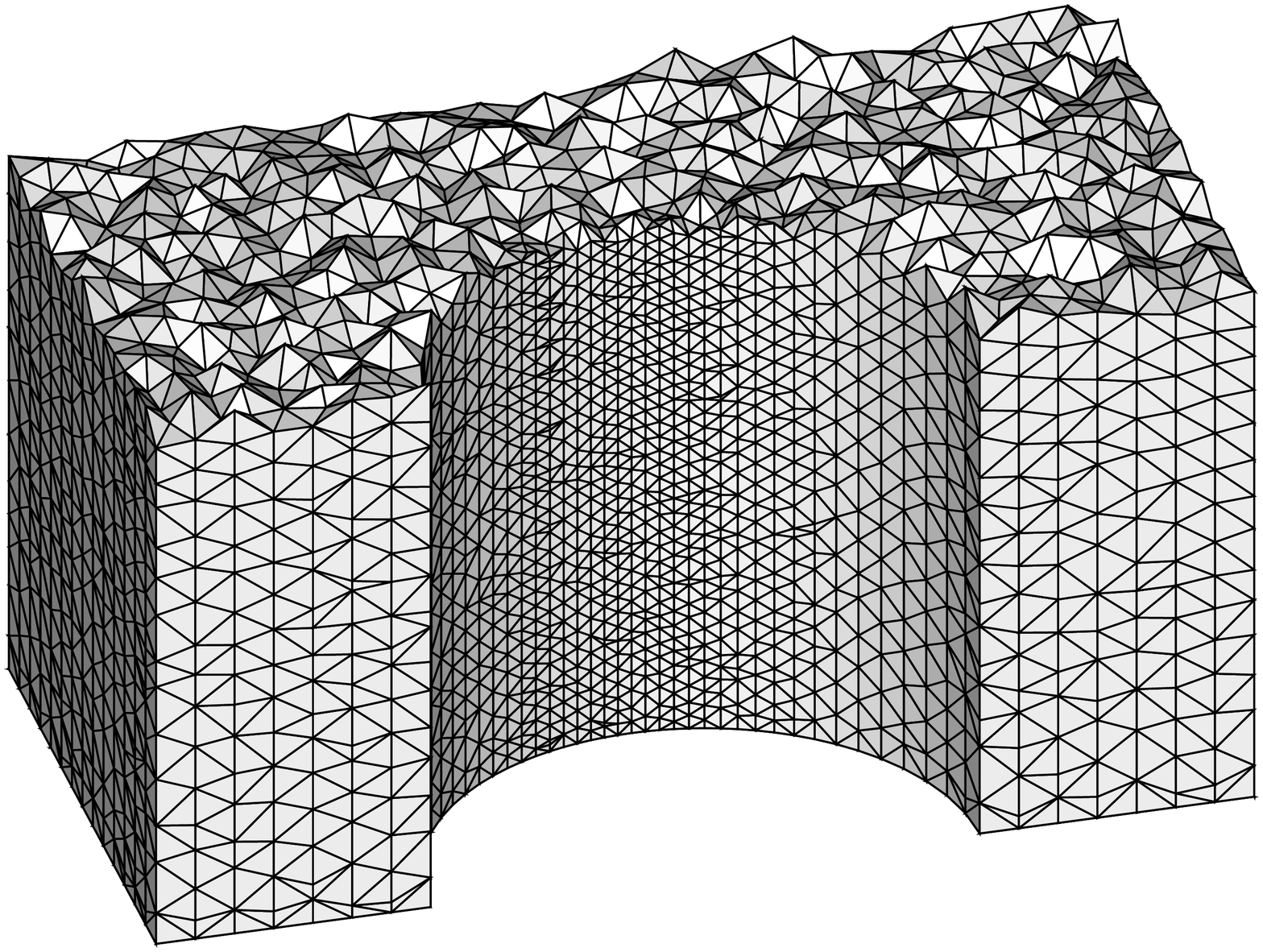} \\ (b)
\end{tabular}
\caption{(a) A typical planar mesh of 1044 triangles.  (b) The
resulting space-time mesh of 55,020 tetrahedra, computed by our
implementation in about 4 seconds.}
\label{Fig:wall}
\end{figure}

\begin{figure}
\centering\footnotesize\sf
\begin{tabular}{@{}c@{}}
\scalebox{1}[-1]{\includegraphics[width=3in]{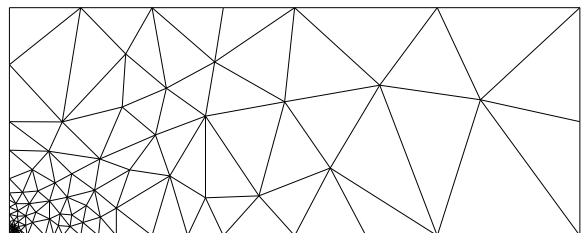}}
\\ (a) \\[4ex]
\includegraphics[width=3in]{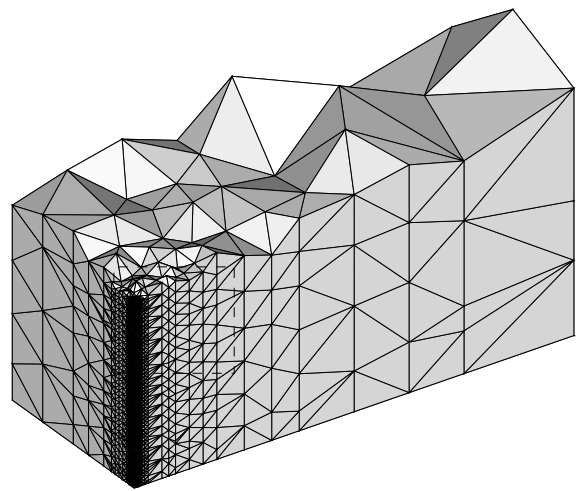}
\\ (b) \\[4ex]
\includegraphics[width=3in]{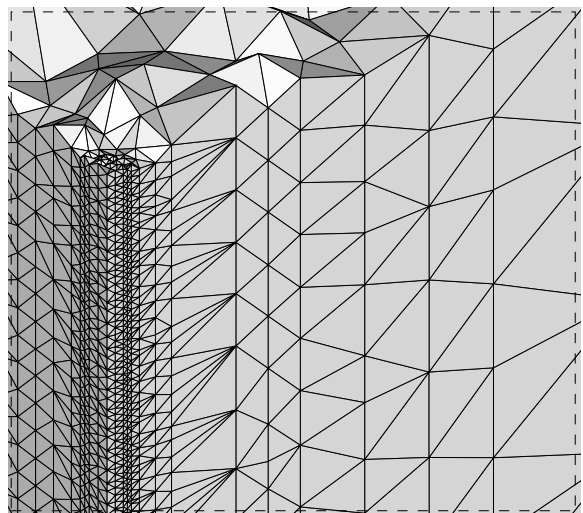}
\\ (c)
\end{tabular}
\caption{(a) A severely graded planar mesh.  (b) The resulting 
space-time mesh.  (c) A close-up of the resulting grading.}
\label{Fig:grade}
\end{figure}

\begin{figure*}[t]
\centering\footnotesize\sf
\begin{tabular}{c@{\qquad}c}
   \scalebox{-1}[1]{\includegraphics[width=2in]{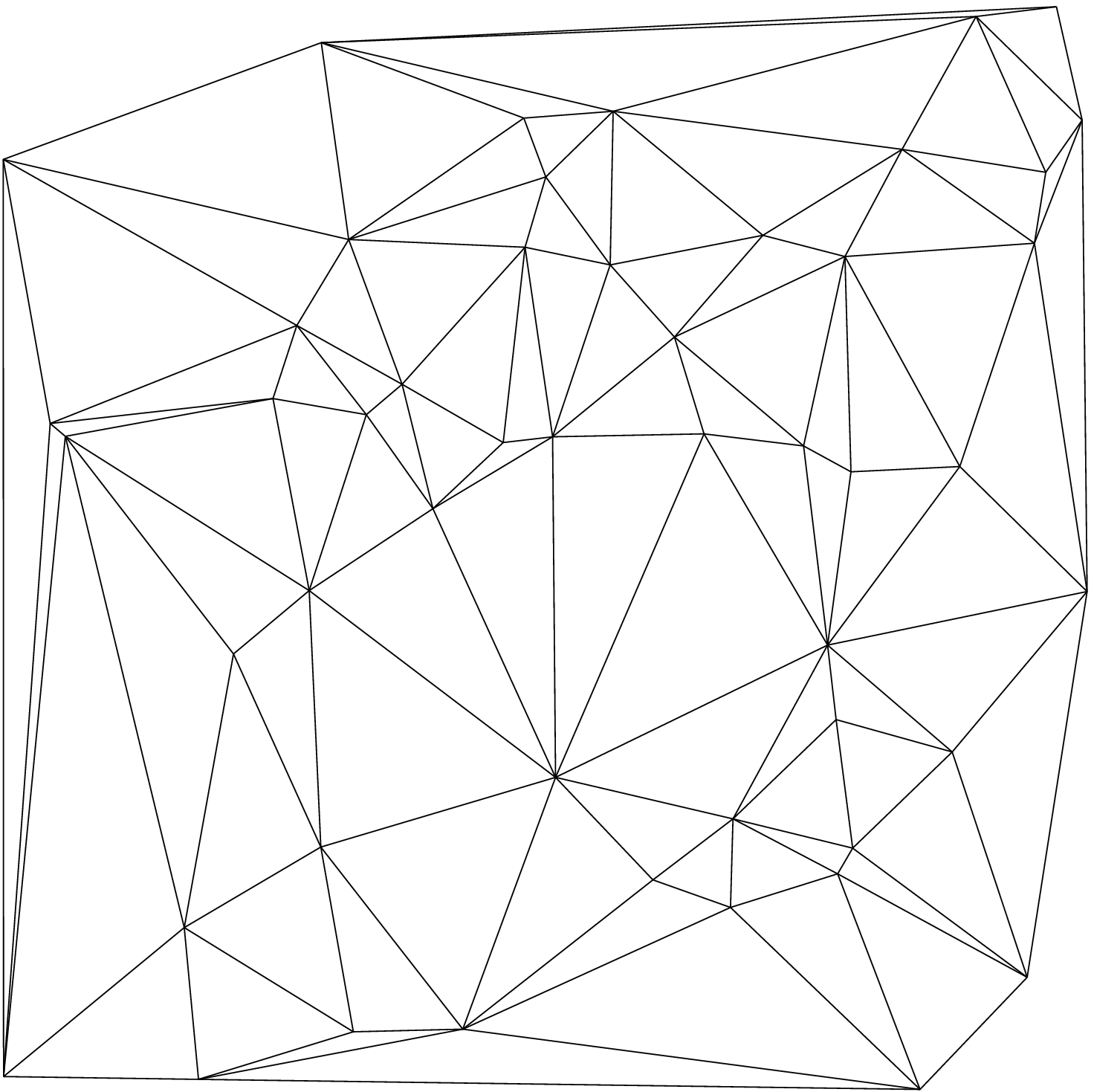}}
&  \scalebox{-1}[1]{\includegraphics[width=2in]{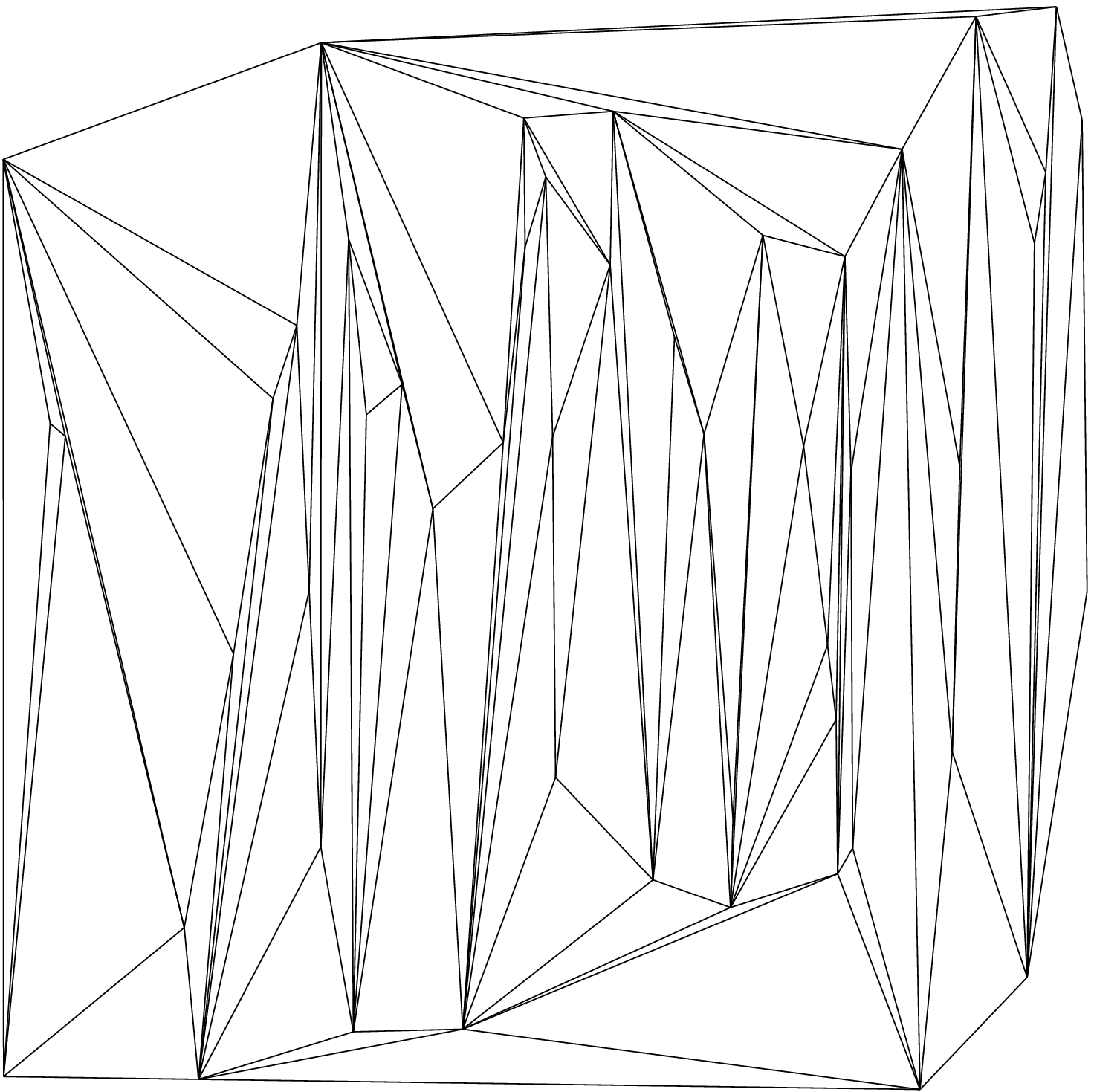}}
\\ (a) & (b)
\\[3ex]
    \includegraphics[width=2.5in]{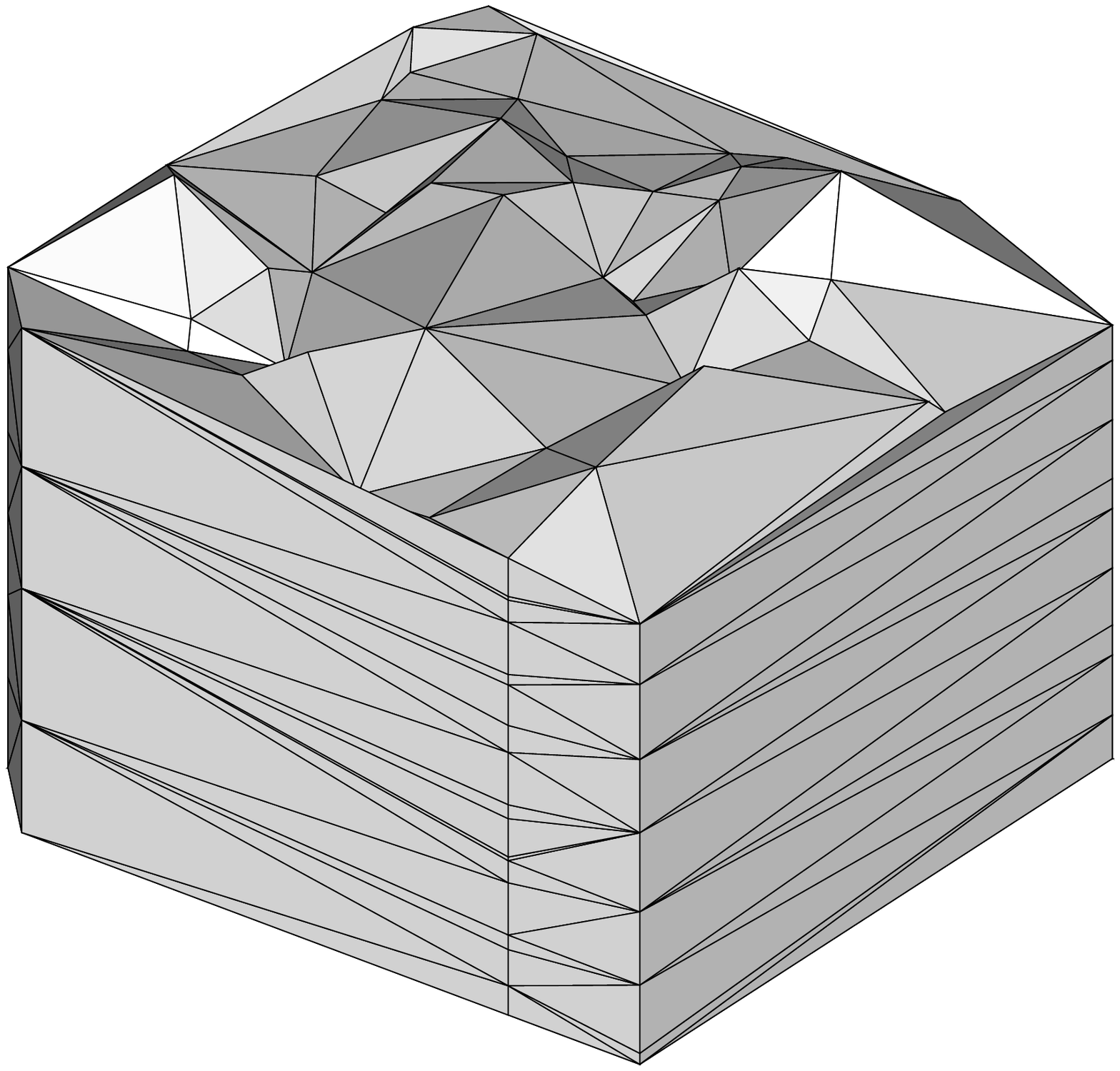}
&   \includegraphics[width=2.5in]{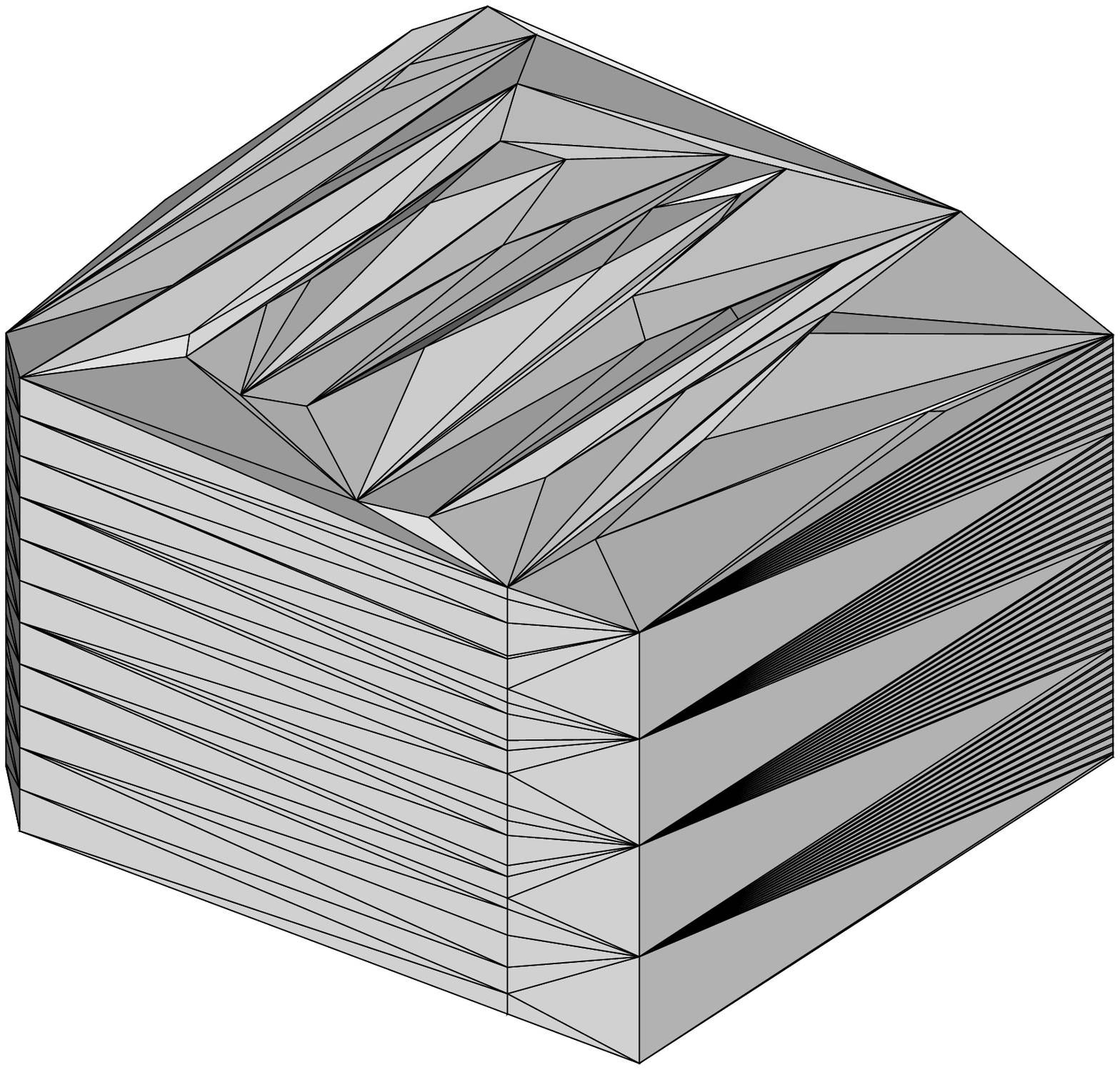}
\\ (c) & (d)
\end{tabular}
\caption{(a) A Delaunay triangulation with a few bad triangles.  (b)
A sweep-line triangulation (of the same point set) with many horrible
triangles.  (c,d) The resulting space-time meshes, showing the 
resulting temporal grading.}
\label{Fig:bad}
\end{figure*}

Figures \ref{Fig:usa} and \ref{Fig:wall}--\ref{Fig:bad} show
space-time meshes computed by our implementation.  In each case, we
stopped advancing each vertex of the front after it passed a target
time value.  In every example, the input triangle mesh contains at
least one (sometimes extremely) obtuse triangle, which caused \Ungor\
and Sheffer's original Tent Pitcher algorithm to fail \cite{UngorS01}.

Our program produces several thousand elements per second, running on
a 1.7~GHz Pentium~IV with 1~gigabyte of memory.  For example, the mesh
in Figure \ref{Fig:usa}, which contains 114,515 tetrahedral elements,
was built from a ground mesh of 2,356 triangles in about 14 seconds.
Figure \ref{Fig:wall} shows an input mesh with 1,044 triangles and the
resulting 55,020-element space-time mesh, which was computed in about
4 seconds.  (These running times include reading and parsing the input
mesh file and writing the output mesh to disk.)

Figure \ref{Fig:grade} illustrates effect of grading in the input mesh
on the size on space-time elements.  The largest and smallest elements
in the ground mesh differ in size by a factor of 128; the resulting
space-time elements differ in duration by a factor of 450.  (The
difference between these two factors might be explained by the obtuse
triangles near the smallest element of the ground mesh.)  Less severe
grading due to varying ground element size can also be seen in Figure
\ref{Fig:wall}.

Figure \ref{Fig:bad} shows the output of our algorithm when the input
mesh is pathological.  The input meshes are the Delaunay triangulation
and a greedy sweep-line triangulation of the same point set.  As
expected, variations in quality in the ground mesh also leads to
temporal grading in our output meshes.  For example, the bottom right
vertex of the space mesh in Figure \ref{Fig:bad}(b) advances much more
quickly than the top right vertex, because it is significantly further
from the lines through any of its neighboring edges.

We tried several different values of the parameter $\e$ in the
progress constraint \eqref{eq:mod}.  All of the example output meshes
were computed using the value $\e \approx 0.1$.  Somewhat to our
surprise, the number of elements in the output mesh varied by only a
few percent as we varied $\e$ from $1/100$ to $1/3$, and smaller
values of~$\e$ usually resulted in meshes with slightly \emph{fewer}
elements, since the modified progress constraint is less severe.
Also, for high-quality ground meshes, where most of the triangles are
acute, the progress constraint affected only a few isolated portions
of the space-time mesh.  On the other hand, smaller values of $\e$
generally led to wider variability in the duration of neighboring
tetrahedra.  As $\e$ increases, the progress guaranteed by Lemma
\ref{L:progress} more closely matches the maximum progress allowed by
the progress constraint; this tends to distribute the progress of each
triangle more evenly among its vertices.


\section{Further Research}
\label{S:outro}

We have presented the first algorithm to generate graded space-time
meshes for arbitrary spatial domains, suitable for efficient use by
space-time discontinuous Galerkin methods.  This is only the first
step toward building a general space-time DG meshing library.

As we mentioned in Section \ref{S:plane}, our algorithm currently
requires the wave speed at any point in space to remain constant or
monotonically decrease over time.  In the short term, we plan to adapt
our algorithm to handle wave speeds that increase over time.  It
should be noted that for many problems, the wave speed is not known in
advance but must be computed on the fly as part of the numerical
solution.

DG methods do not require conforming meshes, where any pair of
adjacent elements meet in a common face.  As a result, fixed time-step
methods allow the space mesh to be refined or coarsened in response to
error estimates, simply by remeshing at any time slice.  Can our
advancing front method be modified to allow for refinement,
coarsening, or other local remeshing operations (like Delaunay flips)?
These operations might be useful not only to avoid numerical error,
but also to make the meshing process itself more efficient.

For many problems, even the boundary of the domain changes over time
according to the underlying system of PDEs.  Can our method be adapted
to handle moving boundaries?  Intuitively, we would like a mesh that
conforms to the boundary as it moves.  This would require us to move
the nodes of the ground mesh continuously over time; remeshing
operations would be required to guarantee that the meshing algorithm
does not get stuck.  Similar issues arise in tracking shocks, which
are surfaces in space-time where the solution changes discontinuously.

Our method currently assumes that all the characteristic cone have
vertical (or at least parallel) axes.  For problems involving fluid
flow, the direction of the cone axis (\ie, the velocity of the
material) varies over space-time as part of the solution.  We could
adapt our method to this setting by overestimating the true tilted
influence cones by larger parallel cones, but intuitively it seems
more efficient to move nodes on the front.  As in the case of moving
boundaries, this would require remeshing the front.  In fact, the
front would no longer necessarily be a monotone polyhedral surface;
extra work may be required to ensure that the resulting mesh is
acyclic.

Finally, to minimize numerical error it is important to generate
space-time meshes of high quality.  Although there are several
possible measures for the quality of a space-time element, further
mathematical analysis of space-time DG methods is required to
determine the most useful quality measures.  This is in stark contrast
to the traditional setting, where appropriate measures of quality and
algorithms to compute high-quality meshes are well known
\cite{BernCER95, BernEG94, ChengDEFT99, Ruppert93, Shewchuk98}. 


\subsection*{Acknowledgments}

The authors thank David Bunde, Michael Garland, Shripad Thite, and
especially Bob Haber for several helpful comments and discussions.

This work was partially supported by NSF ITR grant DMR-0121695.  Jeff
Erickson was also partially supported by a Sloan Fellowship and NSF
CAREER award CCR-0093348.  Damrong Guoy was also partially supported
by DOE grant LLNL B341494.  John Sullivan was also partially supported
by NSF grant DMS-00-71520.  Alper \Ungor\ was also partially supported
by a UIUC Computational Science and Engineering Fellowship.

\hyphenpenalty 0
\bibliographystyle{newabuser}
\bibliography{tezbib}

\end{document}